# Electronic Structure of the Kitaev Material $\alpha$-RuCl$_3$ Probed by Photoemission and Inverse Photoemission Spectroscopies

Soobin Sinn[1,2], Choong Hyun Kim[1,2], Beom Hyun Kim[3,4], Kyung Dong Lee[5], Choong Jae Won[5], Ji Seop Oh[1,2], Moonsup Han[6], Young Jun Chang[6], Namjung Hur[5], Hitoshi Sato[7], Byeong-Gyu Park[8], Changyoung Kim[1,2], Hyeong-Do Kim[1,2,*], and Tae Won Noh[1,2]

[1]Center for Correlated Electron Systems, Institute for Basic Science (IBS), Seoul 08826, Republic of Korea

[2]Department of Physics and Astronomy, Seoul National University (SNU), Seoul 08826, Republic of Korea

[3]Computational Condensed Matter Physics Laboratory, RIKEN, Wako, Saitama 351-0198, Japan

[4]Interdisciplinary Theoretical Science (iTHES) Research Group, RIKEN, Wako, Saitama 351-0198, Japan

[5]Department of Physics, Inha University, Incheon 22212, Republic of Korea

[6]Department of Physics, University of Seoul, Seoul 02504, Republic of Korea

[7]Hiroshima Synchrotron Radiation Center, Hiroshima University, Kagamiyama 2-313, Higashi-Hiroshima 739-0046, Japan

[8]Pohang Accelerator Laboratory, Pohang University of Science and Technology, Pohang 37673, Republic of Korea

*hdkim6612@snu.ac.kr

**Abstract**

Recently, $\alpha$-RuCl$_3$ has attracted much attention as a possible material realization of the honeycomb Kitaev model, which may stabilize a quantum-spin-liquid state. Compared to extensive studies on its magnetic properties, there is still a lack of understanding on its electronic structure, which is strongly related with its Kitaev physics. Here, the electronic structure of $\alpha$-RuCl$_3$ is investigated by photoemission (PE) and inverse photoemission (IPE) spectroscopies. The band gap, directly measured from PE/IPE spectra, is found to be 1.9 eV, much larger than previous estimations. The LDA calculations show that the on-site Coulomb interaction $U$ can open the band gap without spin-orbit coupling (SOC). However, the SOC should also be incorporated to reproduce the proper gap size, indicating that the interplay between $U$ and SOC plays an essential role in the physics of $\alpha$-RuCl$_3$. There exist some spectral features in PE/IPE spectra which cannot be explained by the LDA calculations. To explain such discrepancies, we perform the configuration-interaction calculations for a RuCl$_6^{3-}$ cluster. The experimental data and calculations demonstrate that the 4$d$ compound $\alpha$-RuCl$_3$ is a $J_{\text{eff}}$ = 1/2 Mott insulator rather than a quasimolecular-orbital insulator. Our study also provides important physical parameters, required in verifying the proposed Kitaev physics in $\alpha$-RuCl$_3$.

**Introduction**

The honeycomb Kitaev model has attracted significant attention as a feasible model for a quantum-spin-liquid ground state[1-5]. In this model, a strong spin-orbit coupling (SOC) plays a crucial role, since it provides bond-direction-dependent exchange interaction which results in spin frustration. To realize this possibility, various transition-metal compounds, including $Na_2IrO_3$ and $Li_2IrO_3$ have been investigated[6-8]. These materials contain $5d$ transition metal Ir ions, which have a large SOC strength $\lambda_{SOC} \sim 0.4$ eV[9].

$\alpha$-$RuCl_3$ has recently been added to the list of Kitaev candidates, despite a comparatively modest SOC in the $4d$ Ru ion $\lambda_{SOC} \sim 0.13$ eV[10]. The honeycomb lattice of the system is almost perfect, with the Ru–Cl–Ru angle is close to 90°[11,12]. This make the system ideal for achieving the Kitaev ground state[1-4] even with a relatively small SOC. Numerous experimental studies, including Raman spectroscopy[13,14] and neutron scattering[10], reported that $\alpha$-$RuCl_3$ may be close to the Kitaev-spin-liquid ground state. To distinguish between interesting Kitaev quantum physics and other classical spin fluctuations, it is essential to determine accurate values of the physical parameters related to Kitaev physics, possibly from electronic structure studies.

However, there are still some controversies in the electronic structure of $\alpha$-$RuCl_3$. The magnitude and nature of the band gap remains controversial. An early Hall-effect study of $\alpha$-$RuCl_3$ claimed that the band gap should be about 0.3 eV[15]. Optical studies found the optical gap of 0.3 eV[16], and that was later revised to 1.0 eV[17]. Recently, an angle-resolved photoemission spectroscopy (ARPES) study shows that the Fermi level $E_F$ is located 1.2 eV above the valence band maximum, suggesting that the band gap should be larger than 1.2 eV[18]. There are two possible insulating mechanisms for a spin-orbit coupled $t_{2g}^5$ honeycomb system[19]: A $J_{eff} = 1/2$ Mott insulator[20] and a quasimolecular-orbital band insulator[21]. The model presumes the $J_{\text{eff}} = 1/2$ Mott state[1], but there has been no experimental confirmation for this system. Moreover, the physical parameters characterizing the electronic structure and interactions, which constitute a key input to theoretical descriptions of the unconventional magnetism, have not yet been determined.

Here, we present our experimental and theoretical efforts to understand the electronic structure of $\alpha$-RuCl$_3$ using both photoemission (PE) and inverse photoemission (IPE) spectroscopies. We observed a band gap of about 1.9 eV, much larger than earlier reported values. Local density approximation (LDA) calculations also reveal that the interplay between SOC and electron correlation plays an important role in determining the insulating ground state of $\alpha$-RuCl$_3$. However, some features of the PE/IPE spectra cannot be fully explained by the band calculations, implying a strongly correlated ground state. To explain such detailed features, we performed configuration-interaction (CI) calculations for a RuCl$_6^{3-}$ cluster and determined the microscopic parameters relevant to Kitaev physics.

**Results and Discussion**

The underlying honeycomb symmetry of $\alpha$-RuCl$_3$ can be manifested in constant-energy maps of ARPES data. Figures 1(a)–1(c) shows constant-energy maps at the binding energies of $E_B$ = 1.2, 5.0, and 5.7 eV, respectively. At $E_B$ = 1.2 eV, the crystal symmetry is not clearly resolved. This is probably due to the negligible dispersions of Ru $t_{2g}$ bands. At the higher binding energies, the maps reveal with a six-fold symmetry, which originates from dispersive Cl 3$p$ bands. These constant-energy maps guarantee the high quality of our sample surfaces.

ARPES spectra along the $M\Gamma M$ line in Fig. 1(d) show nearly flat Ru 4$d$ bands near $E_F$ and dispersive Cl 3$p$ bands. Most Ru 4$d$ bands are located between –1.0 to –3.0 eV. The flat and strong peak at –1.5 eV should come from average of many Ru 4$d$ bands. In an enlarged view, the Ru 4$d$ band dispersion is estimated to be about 0.1 eV or less. On the other hand, the Cl 3$p$ bands are located between –3.5 and –7.5 eV, well separated from the Ru 4$d$ bands. Compared to the Ru 4$d$ bands, the Cl 3$p$ bands are highly dispersive. Overall, our ARPES spectra are consistent with recently reported ARPES results[18,22]. Note that the distances between five Ru $t_{2g}$ bands (< 0.2 eV) are much smaller than the band gap (> 1.2 eV). This implies $\alpha$-RuCl$_3$ is not a quasimolecular-orbital insulator, in which the $t_{2g}$ band distances and band gap have common energy scale of $d$-$d$ hopping[21]. For a comparison, we overlaid band dispersions by LDA+SOC+$U$ calculations on the right side of Fig 1(d). The red and blue solid lines correspond to the Ru $t_{2g}$ bands and the Cl 3$p$ bands, respectively. The calculations also support that the flat Ru 4$d$ and the dispersive Cl 3$p$ bands are located near and much below $E_F$, respectively. In spite of this success, there exist some discrepancies between the ARPES spectra and the calculation results, which will be discussed later.

To resolve the controversy on the band gap size of $\alpha$-RuCl$_3$, we exploited combined angle-integrated PE and IPE spectroscopies. Note that the PE/IPE spectra contain information on the DOS of the occupied and unoccupied bands, respectively. Therefore, the combined spectroscopy study of PE and IPE has been established as the most direct method to determine an electronic energy gap[23]. The black dots in Fig. 2 show the both PE and IPE spectra of $\alpha$-RuCl$_3$. The PE spectrum was obtained from the ARPES data of Fig. 1(d) by integrating over momentum. Based on the arguments for Fig. 1(d), we assign the peak around −1.5 eV to Ru $t_{2g}$ antibonding lower Hubbard bands (LHB). The two strong peaks around −4.0 and −7.0 eV should come from the Cl 3$p$ nonbonding and bonding bands, respectively. The right side of Fig. 2 shows an IPE spectrum. We can see prominent two peaks near $E_F$, which are assigned to Ru $t_{2g}$ upper Hubbard bands (UHB) and Ru $e_g$ bands. The crystal-field splitting 10$Dq$, the energy separation between Ru $t_{2g}$ UHB and $e_g$ bands, is estimated to be about 2.2 eV. This value is close to that observed in x-ray absorption spectroscopy[16].

We estimate the band gap of $\alpha$-RuCl$_3$ to be 1.9 eV, which is much larger than those in earlier studies[15-18]. In principle, the size of band gap should correspond to the energy range of zero intensity in PE/IPE spectra. However, the range of zero intensity is reduced due to lifetime and experimental spectral broadening. To overcome such difficulty, we compared the PE/IPE spectra with LDA+SOC+$U$ results. To reproduce the PE/IPE spectra near $E_F$, we used $U - J_H = 4.5$ eV in the calculations. As shown in Fig. 2, the LDA+SOC+$U$ can explain both the valence and conduction bands near $E_F$ reasonably well. Then, the band gap size of $\alpha$-RuCl$_3$ should be around 1.9 eV. Note that this magnitude is definitely higher than 0.3 eV from Hall-effect study[15]. It is also higher than 1.0 eV from recent optical studies, suggesting possible existence of strong exciton effects in the optical spectrum[17].

We found that the interplay between Coulomb interaction $U$ and SOC is essential to understand the physics of $\alpha$-RuCl$_3$. To clarify their roles, we performed LDA calculations with and without $U$ and SOC terms. Figs. 3(a)–3(d) display the results of LDA, LDA+SOC, LDA+$U$, and LDA+SOC+$U$. As shown in Fig. 3(a), without $U$ and SOC, the partially-filled Ru bands with $t_{2g}^5$ electrons should behave as a metal. As shown in Fig. 3(b), the system remains to be a metal with SOC. But, the narrow $t_{2g}$ bands repel each other due to SOC, thus resulting in an apparent total $t_{2g}$ bandwidth broadening as previously reported[24]. On the other hand, as shown in Fig. 3(c), LDA+$U$ results predict a gapped electronic structure, indicating the prime importance of the Coulomb interaction in the insulating nature of $\alpha$-RuCl$_3$. However, the predicted gap size is only about 1.3 eV. Only when we include both SOC and $U$, we can properly describe the observed energy gap value of 1.9 eV. The large enhancement of gap size by 0.6 eV, just by introducing a small SOC of 0.13 eV, indicates that SOC plays a crucial role in the electronic structure of $\alpha$-RuCl$_3$, especially near the Fermi level[25]. Thus, SOC seems to reinforce the correlation strength, or the antiferromagnetic order, by reducing the bandwidth[20,26].

Although the PE/IPE spectra can be explained reasonably well by the LDA+SOC+$U$ calculation results, there are still some discrepancies. In Fig. 1(d) and Fig. 2, there is a sharp nondispersive peak around −2.5 eV, which cannot be explained by the calculations. The orbital character of this peak seems to be Ru 4$d$, because its intensity change is similar to those of the main Ru $t_{2g}$ bands when changing photon energies[22]. Moreover, the clear separation between the Ru 4$d$ and the Cl 3$p$ bands in the ARPES spectrum cannot be reproduced in the calculations. As shown in the IPE spectrum of Fig. 2, size of the crystal-field splitting is also underestimated in the LDA+SOC+$U$ results.

To gain further insights, we carried out CI calculations for a single RuCl$_6^{3-}$ cluster, in which we considered the Ru 4$d$ and the Cl 3$p$ bonding orbitals taking into account of full multiplet structures. Cl nonbonding states around −4 eV are not considered in the calculations to reduce the dimension of the Hilbert space. The relevant Hamiltonian has numerous parameters, including $U$, $J_H$, $\lambda_{SOC}$, 10$Dq$, charge-transfer energy $\Delta$ from Cl 3$p$ to Ru $t_{2g}$ orbitals, and Slater-Koster parameters $t_{pd\sigma}$ and $t_{pd\pi}$. But, many of them can be unambiguously determined, i.e., the $\lambda_{SOC}$ value was adopted from the inelastic neutron scattering study[10], the 10$Dq$ value from the distance between the Ru $t_{2g}$ UHB and the Ru $e_g$ peak, and the $\Delta$ value from the distance between the Cl nonbonding states and the Ru $t_{2g}$ UHB in the PE/IPE spectra. In most transition-metal compounds, $t_{pd\sigma}$ is about twice of $t_{pd\pi}$ [27]. Then, the values of the remaining three parameters $U$, $J_H$, and $t_{pd\pi}$, can be obtained with little errors to fit the band gap, the −2.5 eV peak position, and the position of the $d^5\underline{L}$ final states around −7 eV. The obtained values of parameters are listed in Table 1. Note that the sizes of $U$, $J_H$, and $\Delta$ are difficult to determine without spectroscopic data due to dynamical screening[28,29].

Our CI calculations can explain the spectral features of the PE/IPE data which were difficult to be explained in the LDA+SOC+$U$ calculations. As shown in Fig. 4, the positions of the energy levels from the CI calculations are in good agreement with the peak positions in the PE/IPE spectra. In spite of the moderate SOC, the electronic structure of $\alpha$-RuCl$_3$ is governed by $J_{eff}$ = 1/2 physics, because the electronic energy gap is determined by excited hole and electron states that solely originate from the $J_{eff}$ = 1/2 state. The curious −2.5 eV peak not explained by the band calculations emerges as a high-binding $J_{eff}$ = 1/2 state due to exchange interaction. There exists a clear separation between the Ru 4$d$ and the Cl 3$p$ bands, as in the ARPES spectra. The distance between Ru $t_{2g}$ UHB and $e_g$ peaks, crystal-field splitting, can be also properly explained by our CI calculations. The successful agreements with $J_{eff}$ = 1/2 nature and large $U$ signifies that the 4$d$ compound $\alpha$-RuCl$_3$ has strong local nature and the relativistic Mott ground state instead of the quasimolecular-orbital insulating state[19].

The values of physical parameters, obtained from the CI calculations, are quite necessary for the studies of Kitaev physics in $\alpha$-RuCl$_3$. The strengths of Heisenberg ($J$), Kitaev ($K$), and off-diagonal ($\Gamma$) exchange interactions in the Heisenberg-Kitaev model could be easily obtained[2,3,26] from our values of physical parameters in Table 1. Only one shortcoming is that our CI calculations were performed on a single-site RuCl$_6^{3-}$ cluster, so they did not include direct *d-d* hopping terms between nearest-neighbor Ru ions. To obtain exchange interaction terms, we adopted the values of the *d-d* hopping parameters $t_{dd\sigma}$ and $t_{dd\pi}$ from a recent theoretical study[30]. Then, the exchange strengths of *J, K,* and Γ are determined to −0.7, −1.6, and 1.5 meV, respectively. The magnitudes are much smaller than those by inelastic neutron scattering experiment[10], but are close to those from recent quantum chemistry calculations assuming a *P*3$_1$12 structure[31]. To be more precise, it is highly desirable to perform CI calculations with full Ru 4*d* orbitals[19] for a multi-site cluster, which requires a much larger Hilbert space.

**Conclusion**

In conclusion, we investigated the electronic structure of a Kitaev candidate material $\alpha$-RuCl$_3$. By combining both photoemission and inverse photoemission studies, we directly measured a band gap of 1.9 eV in $\alpha$-RuCl$_3$, which is much larger than the earlier reported values. We also showed that the interplay between electron correlation and spin-orbit coupling plays crucial role in determining the nature of its Mott insulating ground state. Taking into account of the many-body effects using configuration-interaction calculations for a RuCl$_6^{3-}$ cluster, we could obtain the physical parameters and exchange-interaction strengths of the Heisenberg-Kitaev model. The obtained parameters will provide a useful guide to synthesize Kitaev materials with the quantum-spin-liquid state. For example, applying pressure or strain can be a strategy to achieve it. The evolution by the perturbations on Kitaev phase diagram strongly depends on detailed parameters, which we obtained in our study.

**Methods**

**Experiments.** Single-crystalline samples of α-RuCl$_3$ were grown by the self-chemical vapor transport method. Their crystallinity was confirmed by Laue diffraction. All samples were cleaved *in situ* for ARPES and IPE measurements. ARPES measurements were performed at the Beamline 4A1 of Pohang Light Source. ARPES spectra were obtained at a photon energy of $h\nu$ = 70 eV with an energy resolution of 50 meV. During the measurements the sample temperature was kept at 280 K under a vacuum of 3 × 10$^{-11}$ Torr. IPE measurements were carried out at HiSOR[32,33]. Incident electron kinetic energy was set to be 50 eV with an energy resolution of 0.9 eV. The sample temperature was 340 K under a vacuum of 3 × 10$^{-10}$ Torr. IPE spectra were taken in the normal incidence mode. The angular divergence of the electron beam was about 4°, which corresponds to about one-third of the length of the Γ$K$ line. During both ARPES and IPE measurements, we varied photon and electron fluxes but did not observe sample charging effects. The Fermi levels of ARPES and IPE spectra were determined using a gold reference, electrically contacted to the sample.

**Theory.** To calculate the band structure of α-RuCl$_3$, we used local density approximation (LDA) methods. They were calculated by the density functional theory code OPENMX[34] with a zigzag magnetic ordering, which was reported to occur in α-RuCl$_3$[11,35]. In LDA+$U$ and LDA+SOC+$U$ calculations, the $U$ – $J_H$ value of 4.5 eV was employed to reproduce PE/IPE spectra near $E_F$ ($J_H$ is the Hund's coupling.) To explain the fine detailed structures of PE/IPE spectra, we also performed CI calculations[36] for a local RuCl$_6^{3-}$ cluster neglecting nonbonding Cl 3$p$ molecular orbitals. We solved a Hamiltonian for a five-hole system with the Lanczos exact diagonalization method and calculated one-particle Green's functions by spanning eigenvalues of four- and six-hole system with the band Lanczos method[37].

**Acknowledgements**

The authors are grateful to L. J. Sandilands, C. H. Sohn, and S.-H. Baek for invaluable discussions. This work was supported by IBS-R009-D1, by IBS-R009-G2, and by the National Research Foundation of Korea (NRF) grant (2012M2B2A4029470, 2014R1A1A1002868). B. H. Kim is supported by the RIKEN iTHES Project. Experiments at PLS-II were supported in part by MSIP and POSTECH.

**Author contributions**

S.S. and H.D.K. conceived the work. K.D.L., C.J.W., and N.J.H. synthesized $\alpha$-RuCl$_3$ crystals. S.S., J.S.O., H.D.K., and B.G.P. carried out the ARPES experiments, and S.S., J.S.O., H.D.K., and H.S. performed IPE experiments. C.H.K. conducted the LDA calculations, and B.H.K. carried out the CI calculations. The research was carried out with guidance from M.H., Y.J.C., C.K., H.D.K., and T.W.N. All authors discussed the work and commented on the manuscript.

**Table**

| $U$ | $J_\text{H}$ | $\lambda_\text{SOC}$ | $10\,Dq$ | $\Delta$ | $t_{pd\sigma}$ | $t_{pd\pi}$ |
|------|------|------|------|------|------|------|
| 4.35 | 0.35 | 0.13 | 2.2 | 5.0 | 1.90 | -0.90 |

**Table 1. Physical parameters of CI calculations.** The units are in eV. The parameters are determined by reproducing the experimental PE/IPE spectra in Fig. 2 with CI calculations.

**Figures**

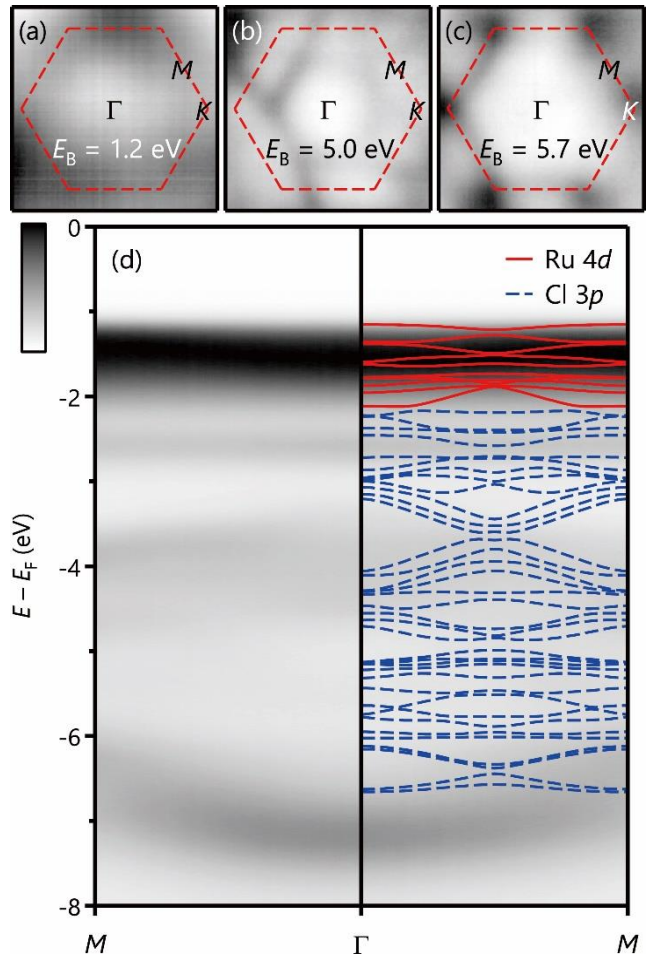

**Figure 1. Momentum-dependent electronic structure of RuCl₃.** ARPES constant-energy maps at different binding energies of (**a**) 1.2 eV, (**b**) 5.0 eV, and (**c**) 5.7 eV. Red dashed hexagon indicates the Brillouin zone of $\alpha$-RuCl$_3$. (**d**) Band dispersions from ARPES along the $M\Gamma M$ line. Calculated bands by LDA+SOC+$U$ ($U - J_H$ = 4.5 eV) are depicted on the right side of (**d**). Red solid lines and blue dashed lines represent Ru 4$d$ and Cl 3$p$ bands, respectively. Note that the existence of a flat band at −2.5 eV and a clear separation between Ru 4$d$ bands and Cl 3$p$ bands are not reproduced in the LDA+SOC+$U$ calculations.

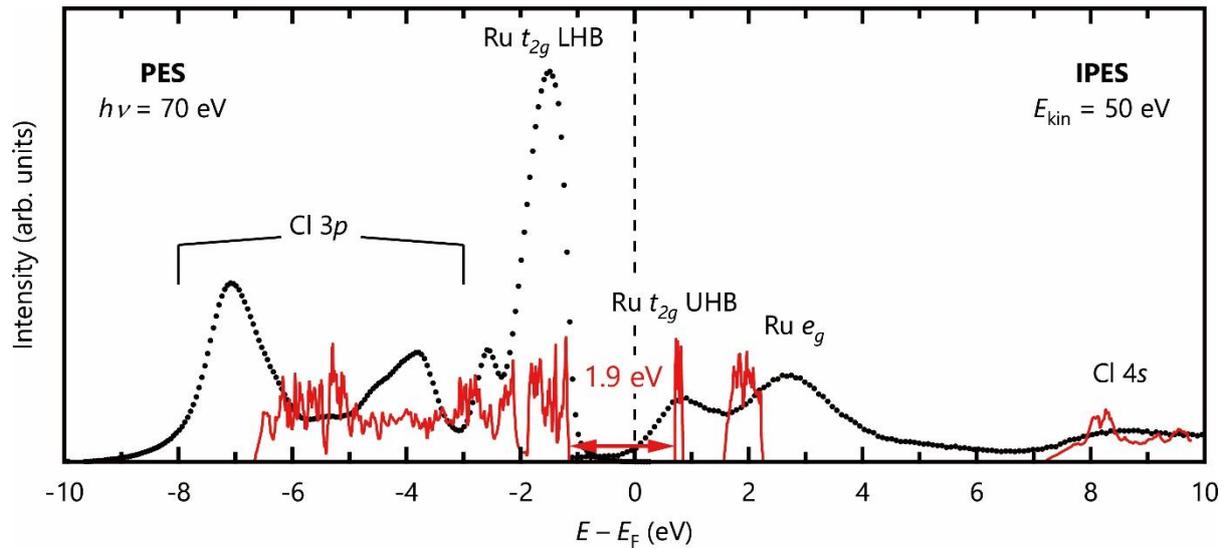

**Figure 2. PE/IPE spectra of α-RuCl₃.** The red solid line represents the density of states from LDA+SOC+$U$ calculations. By comparison between the experiment and the theory, we estimated the size of the band gap to be about 1.9 eV. Note that the crystal field splitting is underestimated in the LDA+SOC+$U$ calculations.

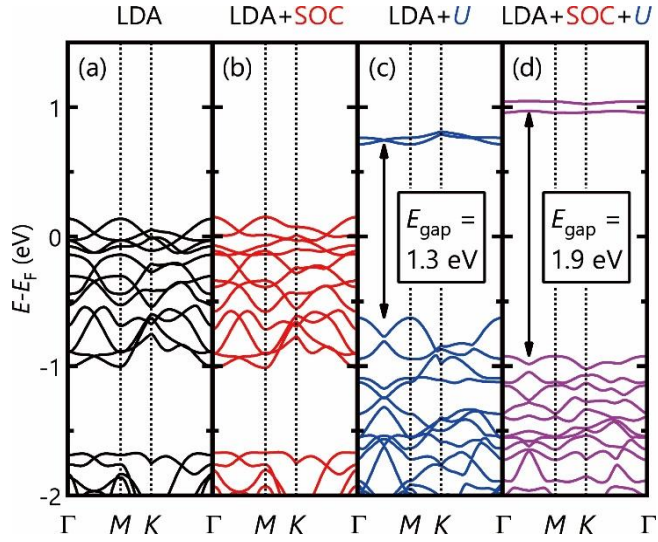

**Figure 3. Electronic band structures of α-RuCl$_3$ by changing *U* and SOC. (a)** LDA, **(b)** LDA+SOC, **(c)** LDA+*U*, and **(d)** LDA+SOC+*U* calculations. Note that the band-gap value about 1.9 eV can be explained only when both *U* and SOC terms are included.

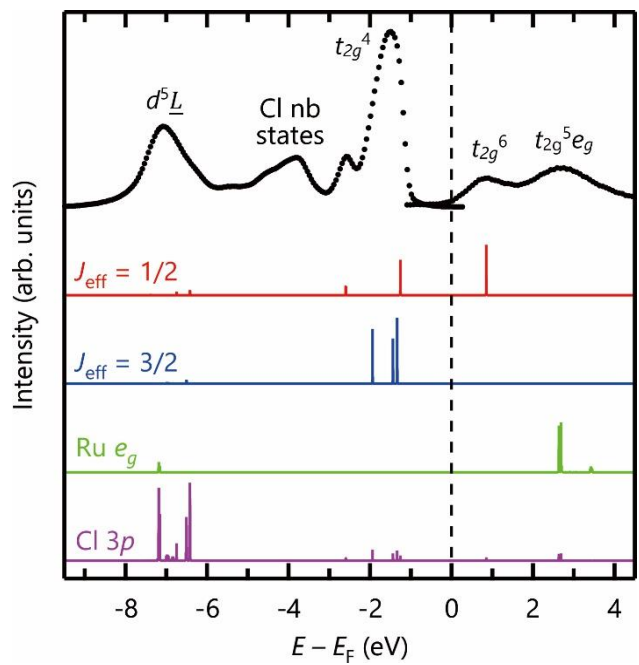

**Figure 4. Comparison of PE/IPE spectra from experiments and CI calculations for a RuCl$_6^{3-}$ cluster.** Spectral weights from CI calculations are shown separately by their spin-orbital characters in the ground state. The electronic energy gap is determined solely by excited states from the $J_{eff} = 1/2$ state. Note that nonbonding Cl 3$p$ orbitals are not included in the calculations, so there is no peak around −4 eV.